\documentclass[a4,11pt]{article}
\usepackage{amssymb, amsmath, latexsym}
\usepackage[margin=1.5cm]{geometry}
\usepackage[titletoc,title]{appendix}
\numberwithin{equation}{section}
\newtheorem{theorem}{Theorem}
\newtheorem{lemma}{Lemma}

\newtheorem{remark}{Remark}
\newtheorem{proof}{Proof}
\title{Cauchy problem on two characteristic hypersurfaces
for the Einstein-Vlasov-Scalar field equations in temporal gauge  }
\author{Marcel Dossa, Jean Baptiste Patenou }
\begin{document}
\maketitle  \abstract{In this paper, we consider the initial value
problem for the Einstein-Vlasov-Scalar field equations in temporal
gauge, where the initial data are prescribed on two characteristic
smooth intersecting hypersurfaces. From a suitable choice of some
free data, the initial data constraints's problem is
 solved globally, then the evolution problem relative to the deduced
initial data is solved locally in time. }
\\ 2010
Mathematics Subject Classification Numbers: 35Q75, 83C05
\section {Introduction}\label{s1}
In this paper, we prove a local (in time) well-posedness result for
the characteristic Cauchy problem for the Einstein-Vlasov-scalar
field equations in temporal gauge, when the data are assigned on two
characteristic intersecting smooth hypersurfaces. Indeed, in General
relativity, there are basically two types of Cauchy problem: the
ordinary spacelike Cauchy problem for the Einstein equations, and
the characteristic initial value problem for these same equations.
In the first case, the system of constraints's equations is
standard, ie. the Hamiltonian and momentum constraints (\cite{m},
Chap. 7; \cite{b}); they depend only of the nature of the Einstein
equations, are independent of the choice of the gauge and of the
corresponding evolution system. In the case of the characteristic
Cauchy problem, the set of constraints's equations includes the
standard constraints and additional constraints induced by the
choice of the gauge, the evolution system considered, the free data,
and obviously the form of the stress-energy momentum tensor of the
considered matter. The gauge commonly used for the characteristic
Cauchy problem is the harmonic gauge, see: \cite{j} where the
Einstein equations in vacuum or perfect fluid are considered, with
data assigned on two null intersecting smooth hypersurfaces,
\cite{d},\cite{e},\cite{g} for the study of the Einstein-Yang-Mills
equations with data prescribed on two intersecting smooth
hypersurfaces, \cite{f} where various aspects of the characteristic
Cauchy problem in General relativity are reviewed,  \cite{l} for the
analysis of the Cauchy problem on a characteristic cone for the
Einstein equations, and \cite{c} where kinematic matter is
considered for the Cauchy problem on a characteristic cone, using
generalized wave map gauge, in the important case of astrophysical
studies.
   Another gauge used recently by other authors is the " Double null foliation gauge" introduced
for the study of the Cauchy problem for the Einstein equations in
vacuum \cite{a},\cite{h}. In presence of the Vlasov's field, the
hierarchical method of resolution of the constraints developed by
A.D. Rendall \cite{j} in the case of the harmonic gauge is less
suitable; this is due to some difficulties caused by the presence of
all the components of the metric in each component of the
stress-energy momentum tensor generated by the Vlasov's matter. Such
difficulties are mentioned in \cite{c},\cite{n},\cite{k}.
Furthermore C. Tadmon in \cite{k} attempted to solve this problem in
harmonic coordinates, but this author was forced to impose unnatural
restrictive conditions (of integral type) for the initial density of
the particles on the initial hypersurfaces. To tackle these
difficulties, we opted for the temporal gauge. Following Y.
Choquet-Bruhat \cite{b},\cite{m}, we chose the evolution system; and
then highlighted the system of constraints which are of two kinds,
the usual Hamiltonian and momentum constraints (\ref{x1}), and other
constraints (\ref{x2}) due to the condition of temporal gauge, all
described in coordinates $(y^\delta)$ (\ref{a2}), adapted to the
geometry of the initial hypersurfaces. The constraints
 $\widetilde{
   \mathfrak{C}}_{ab}-\frac{\widetilde{g}^{cd}\widetilde{\mathfrak{C}}_{cd}}{n-1}\widetilde{g}_{ab}=0,\;(a,b=2,...,n)$,
    extracted from (\ref{x2})
   have their similar in the "Double null foliation gauge" \cite{a}.
    The theorem $1$ of the paper
    resumes in temporal gauge the entire set of the constraints to solve from some free
    data,
    in order to prescribe the full set of initial data for the considered
    evolution system, while theorem $2$ deals with the resolution of these constraints.
    For a suitable choice of some free data, this system
of constraints (\ref{x1})-(\ref{x2}) is solved hierarchically. The
last part of the article is devoted to the existence theorem for the
Einstein Vlasov-Scalar field equations. For sake of simplicity, we
have considered the $\mathcal{C}^\infty$ initial data.
\section{Geometric setting and formulation of the problem}\label{s2}
 Let $(x^\alpha)=(x^0,x^1,x^a),\;(\alpha=0,1,...,n;\;a=2,...,n)$,
denotes the global canonical set of coordinates of
$\mathbb{R}^{n+1}=\mathbb{R}^2\times \mathbb{R}^{n-1},\;(n\geq 3)$.
 $B$ is a compact domain of $\mathbb{R}^{n-1}$,
$Y:=\{(x^\alpha)\in\mathbb{R}^{n+1},\;x^0-|x^1|\geq 0,\;(x^a)\in B
\}$, $\mathcal{I}^0=\{(x^\alpha)\in Y: x^0+x^1=0\}$,
$\mathcal{I}^1=\{(x^\alpha)\in Y: x^0-x^1=0\}$. One considers in
$\widehat{Y}:=Y\times \mathbb{R}^n$ the Cauchy problem for the
Einstein-Vlasov-Scalar field system when the initial hypersurfaces
$\mathcal{I}^0$ and $\mathcal{I}^1$ are null w.r.t. the prescribed
initial data. This system of unknown function $(g,\rho,\Phi)$
reads\begin{eqnarray}\label{x}
  H_g:\; G_{\mu\nu}\equiv R_{\mu\nu}-2^{-1}g_{\mu\nu}R&=& T_{\mu\nu}, \\\label{xx}
  H_\rho:\;p^\alpha\frac{\partial\rho}{\partial x^\alpha}-
    \Gamma_{\mu\nu}^i p^\mu p^\nu\frac{\partial\rho}{\partial p^i}
    &=&0, \\\label{xxx}
  H_\Phi:\; \square\Phi=V'(\Phi).&&
\end{eqnarray}The Einstein equations $H_g$ describe the gravitational potential
$ g$, while the Vlasov equation gives a statistical description of a
collection of particles of rest mass $\textbf{m}$ and density
$\rho\equiv \rho(x,p^0,p^i)$, which move towards the future
$(p^0>0)$ in their mass shell $\mathbb{P}:=\{(x,p^\mu)\in Y\times
\mathbb{R}^{n+1}/g_{\mu\nu}p^\mu p^\nu=-\textbf{m}^2,\;p^0>0\}$. The
wave equation $H_\Phi$ for the matter field $\Phi$ (of potential
$V$) expresses the divergence free of the stress-energy momentum
tensor of matter.
 The terms $R_{\mu\nu},\;R$ and $G_{\mu\nu}$ design
respectively the components of the Ricci tensor, the scalar
curvature, the components of the Einstein tensor $G$, relative to
the searched metric $g$, while the $T_{\mu\nu}$ are the components
of the stress-energy momentum tensor of matter. The
$\Gamma^\lambda_{\mu\nu}$ are the Christoffel symbols of $g$ and the
$p^\lambda$ stand as the components  of the momentum of the
particles w.r.t. the basis $(\frac{\partial}{\partial x^\alpha})$ of
the fiber $\mathbb{P}_x:=\{(p^\alpha)\in
\mathbb{R}^{n+1}/g_{\mu\nu}(x)p^\mu p^\nu=-\textbf{m}^2,\;p^0>0\}$
of $\mathbb{P}$. The $T_{\mu\nu}$ are given by
\begin{equation}\label{e4}
       T_{\alpha\beta}=
     \partial_\alpha \Phi\partial_\beta
  \Phi  -\frac{1}{2}g_{\alpha\beta}(g^{\mu\nu}\partial_\mu\Phi\partial_\nu\Phi +V(\Phi))-
    \int_{\{g(p,p)=-\textbf{m}^2\}}\frac{\rho(x^\nu, p^\mu)p_\alpha
  p_\beta\sqrt{|g|}}{p^0}\;dp^1...dp^n.
    \end{equation}In this setting, we
    choose the temporal gauge \cite{b}, contrary to the
usual harmonic gauge \cite{b}, \cite{c}, \cite{f},\cite{j},
\cite{k}; ie. we choose a zero shift and "densitize" the lapse,
requiring the time to be in wave gauge: $g_{0i}=0,\;\Gamma^0\equiv
g^{\lambda\delta}\Gamma_{\lambda\delta}^0=0,\;i=1,...,n;\lambda,
\delta=0,1,...,n$.
     The metric has then the form
 \begin{equation}\label{e1}
    g=-\tau^2(dx^0)^2+\overline{g}_{ij}dx^idx^j.
\end{equation}
 One denotes by $
\overline{\nabla}$ the connection w.r.t. the induced metric
$\overline{g}$ on $ \Lambda_t: x^0=t$.\\
Setting
$\Lambda_{\mu\nu}=T_{\mu\nu}+\frac{g^{\lambda\delta}T_{\lambda\delta}}{1-n}g_{\mu\nu}$,
and following Choquet Bruhat (see \cite{b},\cite{m}), one chooses as
 evolution system attached to $(H_g,H_\Phi,H_\rho)$, the system
$(H_{\overline{g}},H_\Phi,H_\rho)$ where $H_{\overline{g}}$ is
\begin{equation}\label{x7}
    H_{\overline{g}}:\;\partial_0
R_{ij}-\overline{\nabla}_iR_{j0}-\overline{\nabla}_jR_{i0}=\partial_0
\Lambda_{ij}-\overline{\nabla}_i\Lambda_{j0}-\overline{\nabla}_j\Lambda_{i0};
  \end{equation} and its principal part is $ \square \partial_0 \overline{g}_{ij}$.
The considered problem splits thus in two problems that are the
initial data constraints's problem and the evolution problem for the
third order quasi-diagonal Leray hyperbolic system $
(H_{\overline{g}},H_\Phi,H_\rho)$. The first consists to study how
to prescribe a large class of initial data
$(\overline{g}_0,k_0,\phi_0,\rho_0)$ on
$(\mathcal{I}=\mathcal{I}^0\cup \mathcal{I}^1)\times \mathbb{R}^n$
s.t. for any solution $(\overline{g},\Phi,\rho)$ of the evolution
system $(H_{\overline{g}},H_\Phi,H_\rho)$ in a neighborhood
$\widehat{Y}$ of $\mathcal{I}\times \mathbb{R}^n$ satisfying
$\overline{g}_{|\mathcal{I}}=\overline{g}_0,\;(\partial_0\overline{g})_{|\mathcal{I}}=k_0,\;
\Phi_{|\mathcal{I}}=\phi_0,\;\rho_{|\mathcal{I}}=\rho_0$,
$(g,\Phi,\rho)$ is solution of the Einstein-Vlasov-Scalar field
equations  $ (H_{g},H_\Phi,H_\rho)$ in $\widehat{Y}$, where $g$ is
of the form (\ref{e1}) with $\tau^2=(c(x^i))^2|\overline{g}|$, and
$c$ is determined by the prescribed data such that $\Gamma^0=0$ in
$Y$.
\section{The characteristic initial data constraints}\label{s3}
 To describe the full set of constraints, we introduce the coordinates
$(y^\delta)$ defined by
\begin{equation}\label{a2}
    y^m=x^0+(-1)^m
x^1,\;y^a=x^a; a=2,...,n; m=0,1;
\end{equation}
and require the assumptions:
\begin{equation}\label{m28}
\textbf{(M)}_m:\hbox{ the vector fields
 $\frac{\partial}{\partial y^{1-m}}$ is tangent to the null geodesics generating $\mathcal{I}^m,\; m=0,1$;}
\end{equation}clearly sufficient in temporal gauge to carry out the analysis, and which are similar to the affine
parametrization conditions of \cite{j},\cite{l}. The components of
tensors in coordinates $(y^\delta)$ are equipped with a tilde $"\;\widetilde{}\;"$.\\
The assumptions $\textbf{(M)}_m$ induce that the trace on
$\mathcal{I}^0$ and $\mathcal{I}^1$ of the searched metric $g$ has
the specific form
\begin{equation}\label{m29}
  g_{|\mathcal{I}^m}=\widetilde{g}_{01}(dy^0dy^1+dy^1dy^0)+
    \widetilde{g}_{ab}dy^ady^b;a,b=2,...,n.
\end{equation}Correspondingly, the restrictions to $\mathcal{I}^0$ and
$\mathcal{I}^1$ of the components of the Einstein tensor and the
momentum tensor in coordinates $(y^\delta)$ infer the constraints
(\ref{x2}) below and the following theorem.
 {\theorem{
 Let $(\overline{g},\Phi,\rho)$ be any $\mathcal{C}^\infty$ solution
of the evolution system $(H_{\overline{g}},H_\Phi,\;H_\rho)$
 in a neighborhood $\mathcal{V}$ of $\mathcal{I}\times \mathbb{R}^n$, and let
\begin{eqnarray}\label{a3}
  g&=&-\tau^2(dx^0)^2+\overline{g}_{ij}dx^idx^j;
    \end{eqnarray} s.t. the temporal gauge condition is satisfied in $\mathcal{V}$. One sets
$\widetilde{\mathfrak{C}}_{\mu\nu}=\widetilde{G}_{\mu\nu}-\widetilde{T}_{\mu\nu},\;\widetilde{\mathfrak{C}}=
\widetilde{g}^{\mu\nu}\widetilde{\mathfrak{C}}_{\mu\nu}$ and
     one assumes that w.r.t. the metric $g$, the hypotheses \textbf{$(M_m)$} (\ref{m28}) and the
     relations
\begin{eqnarray}\label{x1}
   \widetilde{
   \mathfrak{C}}_{\overline{m}\lambda}&=&0,
   \lambda=0,...,n;\;m=0,1;\;\overline{m}=1-m;\\\label{x2}
   \;\widetilde{
   \mathfrak{C}}_{ab}-\frac{\widetilde{g}^{cd}\widetilde{\mathfrak{C}}_{cd}}{n-1}\widetilde{g}_{ab}&=&0;
   \widetilde{\mathfrak{C}}_{mm}-2\frac{\widetilde{g}_{01}}{n-1}\widetilde{\mathfrak{C}}+
   \widetilde{g}_{01}\frac{\partial (\widetilde{\Gamma}^0+\widetilde{\Gamma}^1)}{\partial
   y^m}   =0;\;a,b,c,d=2,...,n;
    \end{eqnarray}
are satisfied on $(\mathcal{I}^m),m=0,1$;
     then $(g,\Phi,\rho)$ is a solution of the
    Einstein-Vlasov Scalar field equations $(H_{g},H_\Phi,\;H_\rho)$ in $\mathcal{V}$.
     }\label{th1}}
 {\proof{-}}
 If $(\overline{g},\Phi,\rho)$ is a $\mathcal{C}^\infty$ solution
of the evolution system $(H_{\overline{g}},H_\Phi,\;H_\rho)$ and if
the relations (\ref{a3}), (\ref{x1}),(\ref{x2}) are satisfied for
$g$ of the form \ref{m29}, then on $\mathcal{I}^0$
\begin{equation}\label{a49}
    \mathfrak{C}_{00}=\mathfrak{C}_{01},\;\mathfrak{C}_{11}=\mathfrak{C}_{01},\;
    \mathfrak{C}_{1a}=\mathfrak{C}_{0a},\;\mathfrak{C}_{ab}=-
    \frac{\mathfrak{C}_{01}}{g_{11}}g_{ab},\;
    \mathfrak{\widetilde{C}}=-\frac{n-1}{g_{11}}\mathfrak{C}_{01};
\;a,b=2,...,n;
\end{equation}furthermore, the divergence free properties of the
Einstein tensor $(G_{\mu\nu})$ and the stress energy momentum tensor
of matter $(T_{\mu\nu})$ of $g$ (\ref{a3}), imply that on
$\mathcal{I}^0$ one has
\begin{equation}\label{a50}
    \nabla^\alpha\mathfrak{C}_{\alpha\beta}{_{|\mathcal{I}^0}}=0,\;
\left(\partial_0
               (R_{ij}-\Lambda_{ij})-\overline{\nabla}_{i}\mathfrak{C}_{j0}-
\overline{\nabla}_{j}\mathfrak{C}_{i0}\right){_{|\mathcal{I}^0}}=0.
\end{equation}By combining the relations (\ref{a49}) and (\ref{a50}), one obtains on $\mathcal{I}^0$ the system
\begin{equation}\label{a54}
    g^{11}\partial_1[\mathfrak{C}_{0i}]+L_i([\mathfrak{C}_{0j}])=0;
\end{equation} where for $i=1,...,n$, $L_i$ is
a linear homogeneous expression in terms of
$[\mathfrak{C}_{0j}]\equiv
\mathfrak{C}_{0j}{_{|\mathcal{I}^0}},\;j=1,...,n$. This system has
zero data on $\mathcal{S}$ since the constraints
(\ref{x1})-(\ref{x2}) are satisfied on $\mathcal{S}$, and one
deduces that $[\mathfrak{C}_{0i}]=0$, after that,
$\mathfrak{C}_{\mu\nu}=0$ on $\mathcal{I}^0$ thanks to the relations
(\ref{a49}). In the same way it is shown that
$\mathfrak{C}_{\mu\nu}=0$ on $\mathcal{I}^1$. Now, to prove that
$\partial_0\mathfrak{C}^{\mu\nu}=0$ on $\mathcal{I}^0$, one
restricts to $\mathcal{I}^0$ the following linear homogeneous system
satisfied by the $\mathfrak{C}^{\mu\nu}$ in $\mathcal{V}$ ( see
 \cite{b}, pages [407-414], or
 \cite{i}),
           \begin{eqnarray}\label{e12}
             \partial_0 \mathfrak{C}^{00}+L^{00}(\mathfrak{C}^{\gamma\alpha},\partial_i
\mathfrak{C}^{i0})&=&0 \\\label{e13}
            \partial_0 \mathfrak{C}^{ij}+L^{ij}(\mathfrak{C}^{\gamma\alpha},
\partial_s\mathfrak{C}^{k0})&=&0  \\\label{e14}
   \square_g \mathfrak{C}^{0j}
+L^{0j}(\mathfrak{C}^{\gamma\alpha},\partial_s\mathfrak{C}^{\delta\beta})&=&0.
           \end{eqnarray}By combining these restrictions, the system
(\ref{e14}) restricted to $\mathcal{I}^0$ appears as a homogeneous
linear system of propagation equations on $\mathcal{I}^0$ with
unknowns the restrictions to $\mathcal{I}^0$ of
$\partial_0\mathfrak{C}^{0i},\;i=1,...,n$. To show that this
 system has zero data on $\mathcal{S}$, one restricts to $\mathcal{S}$
the systems (\ref{e12})-(\ref{e13}) and first deduces that on
$\mathcal{S},\;\partial_0 \mathfrak{C}^{00}=0,\;\partial_0
\mathfrak{C}^{ij}=0$, after that, one uses the restrictions to
$\mathcal{S}$ of the properties $\nabla_\alpha\mathfrak{C}^{\alpha
i}=0,\;i=1,...,n$ to conclude that $\partial_0 \mathfrak{C}^{0i}=0$
on $\mathcal{S}$. One deduces that on
$\mathcal{I}^0,\;\partial_0\mathfrak{C}^{0i}=0,\;i=1,...,n$ and
subsequently $\partial_0 \mathfrak{C}^{00}=0,\;\partial_0
\mathfrak{C}^{ij}=0$ on $\mathcal{I}^0$, thanks to the linear system
(\ref{e12})-(\ref{e13}). We remark that similar reasoning holds on
$\mathcal{I}^1$. Finally $\mathfrak{C}^{\mu\nu}=0$ in $\mathcal{V}$
thanks to another linear homogeneous hyperbolic system in
$\mathcal{V}$, derived from (\ref{e12})-(\ref{e14}) which is of
principal part $\square\partial_0 \mathfrak{C}^{\mu\nu}$ ( see
 \cite{b}, pages [407-414], or
 \cite{i}) $\blacksquare$
{\remark{
 We emphasize that the relations (\ref{x1}) and
 (\ref{x2}) depend only of the Cauchy data on $\mathcal{I}^m$ for the evolution system
 $(H_{\overline{g}},H_\Phi,H_\rho)$ as shown in the next
section.}}
\section{The choice of free data and construction of the full set of
initial data for the evolution system
 $(H_{\overline{g}},H_\Phi,H_\rho)$}\label{s4} To proceed to the choice of the free data from
which the full set of initial data of the evolution system
 $(H_{\overline{g}},H_\Phi,H_\rho)$ can be determined, it is necessary to give a
complete description of the constraints (\ref{x1})-(\ref{x2}) in
terms of the Cauchy data of the evolution system
 $(H_{\overline{g}},H_\Phi,H_\rho)$. Indeed, setting
\begin{equation}\label{m34}
    \widetilde{g}_{01}{_{|\mathcal{I}^0}}=\theta,\;\widetilde{g}_{ab}{_{|\mathcal{I}^0}}=\Theta_{ab},\;
    \Phi_{|\mathcal{I}^0}=\phi,\;\rho_{|\mathcal{I}^0}=\textbf{f},\;\psi_{\mu\nu}=
    \frac{\partial\widetilde{g}_{\mu\nu}}{\partial y^0}
,\;\partial_\mu=\frac{\partial}{\partial
y^\mu},\;d\widetilde{p}\;'=d\widetilde{p}^1...d\widetilde{p}^n,
\end{equation}
 the Hamiltonian constraint
$\widetilde{\mathfrak{C}}_{11}=0$ and the momentum constraints
$\widetilde{\mathfrak{C}}_{1a}=0,\;a=2,...,n$, reduce on
$\mathcal{I}^0$ respectively to the following partial differential
relations of the Cauchy data
$(\theta,\Theta_{ab},\;\psi_{11},\;\textbf{f},\;\phi)$ respectively
$(\theta,\Theta_{ab},\;\textbf{f},\;\phi,\;\psi_{1i}),\;i=1,...,n$,
of the evolution system $(H_{\overline{g}},H_\Phi,H_\rho)$, ie.:
\begin{equation*}
    \Theta^{cb}\partial_1\Theta_{cb}\;
\psi_{11}+2\theta\partial_1\left(\Theta^{ab}\partial_1\Theta_{ab}\right)+
\theta\left(\Theta^{cb}\partial_1\Theta_{bd}\right)\left(\Theta^{de}\partial_1\Theta_{ce}\right)
-2\partial_1\theta\Theta^{ab}\partial_1\Theta_{ab}
\end{equation*}
\begin{equation}\label{x8}
= 4\left(\partial_1
\phi\right)^2-2\int_{\mathbb{R}^n}\textbf{f}\;\frac{|\theta|\sqrt{|\Theta|}(\textbf{m}^2+
    \Theta_{ab}\widetilde{p}^a\widetilde{p}^b)^2}{(\widetilde{p}^1)^2}d\widetilde{p}\;'\textbf{;}
\end{equation}
\begin{eqnarray*}
\partial_1\psi_{1a}+\frac{\psi_{11}+\theta
\Theta^{cd}\partial_1\Theta_{cd}-\partial_1\theta}{2\theta}\psi_{1a}+\frac{\psi_{11}
\partial_a
\theta}{2\theta}+\left(\Theta^{cb}\partial_1\Theta_{ca}\right)\partial_b
\theta
-\frac{3}{2}\frac{(\partial_1\theta)(\partial_a\theta)}{\theta}+\partial_{1a}^2\theta&&\\
-\theta \partial_c
(\Theta^{cb}\partial_1\Theta_{ab})+\frac{1}{2}\partial_a
(\Theta^{cd}\Theta_{cd})-\theta^2\partial_a\psi_{11}-
\Theta^{cb}(\partial_c\theta)(\partial_1\Theta_{ba})-
\frac{\theta}{2}\Theta^{de}\Theta^{cb}(\partial_1\Theta_{ba})(\partial_c\Theta_{de})&&
\end{eqnarray*}
\begin{equation}\label{x9}
+\frac{\theta}{2}\Theta^{de}\Theta^{cb}(\partial_1\Theta_{db})(\partial_a\Theta_{ec}+\partial_c\Theta_{ea}-
\partial_e\Theta_{ac})=
2\theta \left(\partial_1 \phi\partial_a
   \phi
   +\int_{\mathbb{R}^n}\frac{\textbf{f}|\theta|\sqrt{|\Theta|}(\textbf{m}^2+\Theta_{cd}
    \widetilde{p}^c\widetilde{p}^d)\Theta_{ab}\widetilde{p}^b}
    {(\widetilde{p}^1)^2}d\widetilde{p}\;'\right)\textbf{.}
\end{equation}
 The constraint $\mathfrak{\widetilde{C}}_{10}=0$ is in turn a
partial differential relation of the Cauchy data
$(\theta,\Theta_{ab},\;\textbf{f},\;\phi,\;\psi_{1i},\;\psi_{ab}),\;i=1,...,n,\;a,b=2,...,n$,
of the evolution system $(H_{\overline{g}},H_\Phi,H_\rho)$ which, by
setting $\chi=\Theta^{ab}\psi_{ab}$, reads
\begin{eqnarray*}
  \partial_1\chi +\frac{1}{2}(\Theta^{cd}\partial_1\Theta_{cd}-\frac{\psi_{11}}{\theta})\chi -
  \Theta^{ab}\partial_a\psi_{1b}+\frac{1}{2}\Theta^{cd}\Theta^{ab} (2\partial_a\Theta_{cb}-
  \partial_c\Theta_{ab})\psi_{1d}+\theta\Theta^{ab}\partial^2_{ab} \ln |\theta|+&&  \\
  \frac{\theta}{2}\Theta^{ab}\partial_b (\Theta^{cd}\partial_a \Theta_{cd})-
  \frac{\theta}{2}\Theta^{ab} \partial_c\left[\Theta^{cd}(2\partial_a\Theta_{bd}-\partial_d\Theta_{ab})\right]+
  \frac{1}{2\theta}\Theta^{ab}(\psi_{1a}\psi_{1b}+\partial_a\theta\partial_b\theta)+
  \theta\Theta^{ab}\widetilde{\Gamma}_{ac}^d\widetilde{\Gamma}_{db}^c &&
\end{eqnarray*}
\begin{equation}\label{x10}
    -\theta(\partial_c\ln
    |\theta|+\widetilde{\Gamma}^d_{dc})\Theta^{ab}\widetilde{\Gamma}^c_{ab}=
    \left(\Theta^{cd}\partial_c \phi
\partial_d\phi+V(\phi)\right)
+2\int_{\mathbb{R}^n}\textbf{f}\;\frac{
  \theta\sqrt{|\Theta|}(\textbf{m}^2+
  \Theta_{ab}\widetilde{p}^a\widetilde{p}^b)
}{\widetilde{p}^1}\;d\widetilde{p}\;';
\end{equation}with
 $   \widetilde{\Gamma}^c_{ab}=\frac{1}{2}\Theta^{cd}(\partial_a\Theta_{bd}+
   \partial_b\Theta_{ad}-\partial_d\Theta_{ab}),\;
    a,b,c,d=2,...,n$.\\
For the constraints's equations $
  \mathfrak{\widetilde{C}}_{ab}-\frac{\widetilde{g}^{cd}
\displaystyle\mathfrak{\widetilde{C}}_{cd}}{n-1}\widetilde{g}_{ab}=0
,
  (a,b,c,d=2,...,n)$, extracted from (\ref{x2}), they are also partial differential
  relations of the Cauchy data
  $(\theta,\Theta_{ab},\;\textbf{f},\;\phi,\;\psi_{1i},\;\psi_{ab}),\;(i=1,...,n,\;a,b=2,...,n)$,
of the system $(H_{\overline{g}},H_\Phi,H_\rho)$, ie.:
    \begin{eqnarray*}
     \partial_1\psi_{ab}+\frac{1}{2}\left((-\frac{\psi_{11}}{\theta}+\Theta^{ef}\partial_1\Theta_{ef})
     \delta_a^c\delta_b^d-\frac{1}{\theta}(\Theta^{ce}\partial_1\Theta_{ea}\delta_b^d+
     \Theta^{ed}\partial_1\Theta_{eb}\delta_a^c)\right)\psi_{cd}
       &&  \\
    -\frac{\theta}{2}
     \partial_c\left(\Theta^{cd}(\partial_b\Theta_{da}+\partial_a\Theta_{db}-\partial_d\Theta_{ab})\right)+
     \frac{1}{2}\Theta^{cd}\psi_{1d}(\partial_b\Theta_{ca}+\partial_a\Theta_{cb}-\partial_c\Theta_{ba})
      &&\\
     -\frac{1}{2}(\partial_b\psi_{1a}-\partial_a\psi_{1b})+\theta\partial^2_{ab}\ln
     |\theta|+\frac{\theta}{2}\partial_b(\Theta^{cd}\partial_a\Theta_{cd})+
     \frac{1}{4\theta}\psi_{1a}(2\psi_{1b}-\partial_b\theta)+&&
   \end{eqnarray*}
   \begin{equation}\label{x11}
     \widetilde{\Gamma}_{ac}^d
     \widetilde{\Gamma}^c_{db}+\frac{\theta}{2}\partial_1\Theta_{ab}\chi -\theta (\partial_c\ln |\theta|+
     \widetilde{\Gamma}^d_{dc})\widetilde{\Gamma}^c_{ab}=
     \frac{\widetilde{R}^{(n-1)}-\Theta^{cd}\widetilde{T}_{cd}}{n-1}\Theta_{ab}+\widetilde{T}_{ab};\;
     a,b,c,d,e,f=2,...,n;
   \end{equation}
      \begin{eqnarray*}
 &&\widetilde{R}^{(n-1)}\equiv \Theta^{cd}\widetilde{R}_{cd}
 = \frac{\partial_1\chi}{\theta} +
 \frac{1}{2\theta}(\Theta^{cd}\partial_1\Theta_{cd}-\frac{\psi_{11}}{\theta})\chi -
 \frac{ \Theta^{ab}\partial_a\psi_{1b}}{\theta}+ \Theta^{ab}\partial^2_{ab} \ln |\theta|+ \\
  &&\frac{1}{2\theta}\Theta^{cd}\Theta^{ab} (2\partial_a\Theta_{cb}-
  \partial_c\Theta_{ab})\psi_{1d}+
  \frac{1}{2}\Theta^{ab}\partial_b (\Theta^{cd}\partial_a
  \Theta_{cd})-\frac{1}{2}\Theta^{ab}
  \Theta^{cd}(2\partial_a\Theta_{bd}-\partial_d\Theta_{ab})+
  \end{eqnarray*}
\begin{equation}\label{a17}
  \frac{1}{2\theta^2}\Theta^{ab}(\psi_{1a}\psi_{1b}+\partial_a\theta\partial_b\theta)+
  \Theta^{ab}\widetilde{\Gamma}_{ac}^d\widetilde{\Gamma}_{db}^c   -(\partial_c\ln
    |\theta|+\widetilde{\Gamma}^d_{dc})\Theta^{ab}\widetilde{\Gamma}^c_{ab};
\end{equation}
   \begin{equation}\label{a31}
\widetilde{T}_{ab} = \partial_a \phi
  \partial_b \phi-\frac{1}{2}\Theta_{ab}\left(\frac{2}{\theta}\left[\frac{\partial \Phi}{\partial
y^0}\right]\left( \partial_1\phi\right)+\Theta^{cd}\partial_c
\phi\partial_d\phi+V(\phi)\right)-\int_{\mathbb{R}^n}\textbf{f}\;\frac{2\theta\sqrt{|\Theta|}
  \Theta_{ae}\Theta_{bf}\widetilde{p}^e\widetilde{p}^f }{\widetilde{p}^1}
  \;d\widetilde{p}'.
\end{equation}
  The last constraint
    $\mathfrak{\widetilde{C}}_{00}-2\frac{\widetilde{g}_{01}}{n-1}\mathfrak{\widetilde{C}}+\widetilde{g}_{01}
    \frac{\partial (\widetilde{\Gamma}^0+\widetilde{\Gamma}^1)}{\partial
    y^0}=0$ is a partial differential relation comprising all the Cauchy
data of the evolution system $(H_{\overline{g}},H_\Phi,H_\rho)$,
ie.:
\begin{eqnarray*}
     \partial_1\psi_{01}-\frac{1}{2}(\frac{\chi}{2}+\partial_1\ln |\theta|)\psi_{01}+
    \frac{\theta}{2}\partial_1\chi -\frac{3\theta}{2}\Theta^{cd}\psi_{1d}\partial_c\theta +
    \frac{\theta}{2}(\partial_d\Theta^{dc})\partial_c\theta +
    \frac{\theta}{2}\Theta^{cd}\partial_{cd}^2\theta&& \\
     -\frac{3}{4}\Theta^{cb}\psi_{1b}\psi_{1c}+\frac{1}{4}\Theta^{cb}\partial_b\theta\partial_c\theta
     -\frac{\chi}{4}\psi_{11}+\frac{\theta}{2}\Theta^{cb}\widetilde{\Gamma}^d_{cd}\partial_b\theta
     -\frac{1}{2}(\partial_1\ln|\theta|)\psi_{11}-\frac{1}{8}(\Theta^{cb}\partial_1\Theta_{cb})
     \psi_{11}&&  \\
      +\frac{\partial_1\psi_{11}}{2}-\frac{\psi_{11}^2}{2\theta}-
      \frac{\theta}{4}\Theta^{ab}\partial_1\psi_{ab}-\frac{\psi_{11}\partial_1\theta}{2\theta}-
      \Theta^{ad}\psi_{1d}\psi_{1a}+\frac{\theta}{2}\Theta^{ac}\Theta^{bd}(\partial_1\Theta_{ab})\psi_{cd}&&
   \end{eqnarray*}
   \begin{equation}\label{x12}
   =\frac{\theta}{2}\left(\left[\frac{\partial \Phi}{\partial
  y^0}\right]^2-2\int_{\mathbb{R}^n}\textbf{f}\;|\theta|^3 \sqrt{|\Theta|}
  \;\widetilde{p}^1\;d\widetilde{p}'\right)-\frac{\theta^2}{2}
        (\widetilde{R}^{(n-1)}+2\frac{\Theta^{ab}\widetilde{T}_{ab}}{n-1}).
   \end{equation}
We note that a similar description of the constraints
(\ref{x1})-(\ref{x2}) ( for $m=1$)
   in terms of the Cauchy data
$(\underline{\theta},\;\underline{\Theta}_{ab}
,\;\underline{\psi}_{1\nu},\;\underline{\psi}_{ab},\;\underline{\phi},\;\underline{\textbf{f}})$
on $\mathcal{I}^1\times \mathbb{R}^n$ of the evolution system
$(H_{\overline{g}},H_\Phi,H_\rho)$ is also valid, where:
\begin{equation}\label{m35}
    \underline{\theta}:=\widetilde{g}_{01}{_{|\mathcal{I}^1}},\;\underline{\Theta}_{ab}:=
    \widetilde{g}_{ab}{_{|\mathcal{I}^1}}
    ,\;\underline{\phi}:=
    \Phi_{|\mathcal{I}^1},\;\textbf{\underline{f}}:=\rho_{|\mathcal{I}^1},\;\underline{\psi}_{\mu\nu}:=
    \frac{\partial\widetilde{g}_{\mu\nu}}{\partial y^1}.
\end{equation}
 \textbf{The free data}
  The free data making possible the resolution of the
constraints (\ref{x1})-(\ref{x2}) comprise:\\
(\textbf{a})- $\mathcal{C}^\infty$ functions
$\widetilde{\gamma}_{ab}=\gamma_{ab}(y^0-y^1,y^a)$ where the
$\gamma_{ab}\equiv \gamma_{ab}(x^1,x^a)$ are $\mathcal{C}^\infty$
functions of the variables $x^1,x^a$ that make up a symmetric
positive definite matrix satisfying $\left|\gamma^{ab}\frac{\partial
\gamma_{ab}}{\partial
x^1}\right|>0$;\\
 (\textbf{b})- smooth functions
$(\theta,\;\phi)$ on $\mathcal{I}^0$, and $\textbf{f}$ on
$\mathcal{I}^0\times \mathbb{R}^n$ (respectively
 $(\underline{\theta}
,\;\underline{\phi})$ on $\mathcal{I}^1$, and
$\textbf{\underline{f}}$ on $\mathcal{I}^1\times \mathbb{R}^n$) s.t.
$\theta,\;\underline{\theta}$ are negative, $\textbf{f}$
(respectively $\textbf{\underline{f}}$) is non negative of compact
support contained in $\{\widetilde{p}^1>c_1>0\}$ (respectively
$\{\widetilde{p}^0>c_0>0\}$) for a mass $\textbf{m}\neq 0$; and for
the zero mass the support of $\textbf{f}$ (respectively
$\textbf{\underline{f}}$) is contained in
$\{\widetilde{p}^1>c_1>0,\;\sum_{a=2}^n(\widetilde{p}^a)^2>c'_2>0\}$
(respectively
$\{\widetilde{p}^0>c_0>0,\;\sum_{a=2}^n(\widetilde{p}^a)^2>c_2>0\}$),
besides that, $ Supp(\textbf{f})\cap (\mathcal{S}\times
\mathbb{R}^n)=\varnothing$, $Supp(\underline{\textbf{f}})\cap
(\mathcal{S}\times \mathbb{R}^n)=\varnothing $; and one has
compatibilities relations
\begin{equation}\label{m37}
    \theta=\underline{\theta},\;\phi=\underline{\phi},\;\hbox{on
$\mathcal{S}$ },\;\textbf{f}=\textbf{\underline{f}}\hbox{ on
$\mathcal{S}\times \mathbb{R}^n$}.
\end{equation}
 {\theorem{Given the free data as described above by
(\textbf{a})-(\textbf{b}). Then, there exists a unique global
solution $(\theta,\;\Theta_{ab}
,\;\psi_{1\nu},\;\psi_{ab},\;\phi,\;\textbf{f})$ on
$\mathcal{I}^0\times \mathbb{R}^n$ and
$(\underline{\theta},\;\underline{\Theta}_{ab}
,\;\underline{\psi}_{1\nu},\;\underline{\psi}_{ab},\;\underline{\phi},\;\underline{\textbf{f}})$
on $\mathcal{I}^1\times \mathbb{R}^n$) of the initial data
constraints (\ref{x1})-(\ref{x2}) for the Einstein-Vlasov Scalar
field equations.}\label{th2}} {\proof{We concentrate on the case of
$\mathcal{I}^0\times \mathbb{R}^n$ and an analogous scheme holds on $\mathcal{I}^1\times \mathbb{R}^n$ }}\\
Given the free data (\textbf{a})-(\textbf{b}), one solves the
constraints . Indeed, for the case of $\mathcal{I}^0$, let set
$\Theta_{ab}(y^1,y^a)=\gamma_{ab}(-y^1,y^a),\;\underline{\Theta}_{ab}(y^0,y^a)=\gamma_{ab}(y^0,y^a)$;
then $|\Theta^{ab}\partial_1\Theta_{ab}|>0$, and $\psi_{11}$ solves
algebraically the Hamiltonian constraint
$\mathfrak{\widetilde{C}}_{11}=0$ as described by (\ref{x8}). The
other constraints (\ref{x9})-(\ref{x12}) are hierarchical linear
ordinary differential equations of the variable $y^1$ depending
smoothly on the parameters $y^a,\;a=2,...,n$. They are solved
hierarchically via the theory of ordinary differential systems,
using the initial conditions
\begin{equation}\label{a1}
 {\psi_{ab}}_{|\mathcal{S}}=
    \frac{\partial \gamma_{ab}}{\partial
    x^1}_{|\mathcal{S}},\;
    {\underline{\psi}_{ab}}_{|\mathcal{S}}=-
    \frac{\partial  \gamma_{ab}}{\partial x^1}_{|\mathcal{S}},
{\psi_{01}}_{|\mathcal{S}}=
    \frac{\partial \underline{\theta}}{\partial y^0}_{|\mathcal{S}}
    ,\;{\underline{\psi}_{01}}_{|\mathcal{S}}=
    \frac{\partial \theta}{\partial
    y^1}_{|\mathcal{S}},
    {\psi_{1i}}_{|\mathcal{S}}=0,\;\underline{\psi}_{0i}{_{|\mathcal{S}}}=0;
    \end{equation}  one obtains
a unique global solution
$(\psi_{1a},\;\psi_{ab},\;\psi_{01}),a,b=2,...,n$ . Concretely, one
considers the momentum constraint $\widetilde{\mathfrak{C}}_{01}=0$
described in (\ref{x10}) and the constraints
$Z_{ab}\equiv\mathfrak{\widetilde{C}}_{ab}-\frac{\widetilde{g}^{cd}
\displaystyle\mathfrak{\widetilde{C}}_{cd}}{n-1}\widetilde{g}_{ab}=0$
described in (\ref{x11}) for $(a,b)\neq (2,2)$ since
$(Z_{ab}),\;(a,b=2,...,n)$, is a traceless tensor. One can first
solve the constraint (\ref{x10}) of unknown
$\chi=\Theta^{ab}\psi_{ab}$. After that one solves the constraints
(\ref{x11}) of unknowns $\psi_{ab}$ for $(a,b)\neq
    (2,2)$ provided $\psi_{22}$ takes the value $
\psi_{22}=\frac{1}{\Theta^{22}}(\chi-\displaystyle\sum_{(a,b)\neq
    (2,2)}\Theta^{ab}\psi_{ab})$, and where $\Theta^{ab}\widetilde{R}_{ab}\equiv
\widetilde{R}^{(n-1)}$ equals $-2\frac{\widetilde{T}_{01}}{\theta}$
since $\mathfrak{\widetilde{C}}_{01}=0$ is satisfied.
 That $Z_{22}=0$ is also satisfied with $
\psi_{22}=\frac{1}{\Theta^{22}}(\chi-\displaystyle\sum_{(a,b)\neq
    (2,2)}\Theta^{ab}\psi_{ab})$ follows
from the traceless property of $(Z_{ab})$. For the outgoing
derivative $\left[\frac{\partial\Phi}{\partial y^0}\right]$ which
appears in the constraints (\ref{x11}), (\ref{x12}), it is obtained
by solving, using the initial datum
$\left[\frac{\partial\Phi}{\partial
y^0}\right]_{|\mathcal{S}}=\frac{\partial \underline{\phi}}{\partial
y^0}(0,y^a)$, the propagation equation obtained by restricting the
equation $H_\Phi$ to $\mathcal{I}^0$. At least the constraint
(\ref{x12}) determines $\psi_{01}\blacksquare$
\section{Resolution of the evolution system
$(H_{\overline{g}},H_\Phi,H_\rho)$}\label{s5} {\theorem{Given the
free data $ \gamma=(\gamma_{ab})$ on $\mathbb{R}^n$,
 $(\theta,\;\phi)$ on $\mathcal{I}^0$, and $\textbf{f}$ on
$\mathcal{I}^0\times \mathbb{R}^n$ (respectively
 $(\underline{\theta}
,\;\underline{\phi})$ on $\mathcal{I}^1$, and
$\textbf{\underline{f}}$ on $\mathcal{I}^1\times \mathbb{R}^n$); as
described in section \ref{s4}.
 Then, there exists a unique (4-tuple) $(\mathcal{V},g,
     \Phi,\rho)$ s.t.:
      $\mathcal{V}$ is a neighborhood of $\mathcal{S}$
     in $\mathcal{Y}:=\{(y^\mu)\in \mathbb{R}^{n+1}/y^0\geq 0,\;y^1\geq 0\}$;
     $g$ is a Lorentzian metric on
     $\mathcal{V}$ of the form
     \begin{equation}
        g=-\frac{|\overline{g}|}{|\gamma|}(dx^0)^2+g_{ij}dx^idx^j;\;\overline{g}=(g_{ij});
     \end{equation} $(g,\Phi,\rho)$ is a $\mathcal{C}^\infty$
     solution of the
     Einstein-Vlasov-Scalar field equations in $\mathcal{V}\times
      \{(p^\mu)\in \mathbb{R}^{n+1}/g_{\mu\nu}p^\mu
      p^\nu=-\textbf{m}^2,\;p^0>0\}$ with $\rho$ of compact support; and
 $
{\widetilde{g}_{01}}{|_{\mathcal{I}^0}}=\theta,\;{\widetilde{g}_{01}}{|_{\mathcal{I}^1}}=\underline{\theta},
{\widetilde{g}_{ab}}{|_{\mathcal{I}^0}}={\gamma_{ab}(-y^1,y^a)},
{\widetilde{g}_{ab}}{|_{\mathcal{I}^1}}={\gamma_{ab}(y^0,y^a)}
,\;\Phi_{|\mathcal{I}^0}=\phi,
\Phi_{|\mathcal{I}^1}=\underline{\phi},\;\rho_{|(\mathcal{I}^0\times
\mathbb{P}_x)}=\textbf{f}, \rho_{|(\mathcal{I}^1\times
\mathbb{P}_x)}=\textbf{\underline{f}}$.
             }\label{th3}}
{\proof{Sketch of the proof (See Appendix \ref{ap} for a more
extended
version the proof)}}\\
Let denote by $(\overline{g}_0,k_0,\rho_0,\phi_0)$ the solution of
the initial data constraints problem as constructed in section 4,
from the free data. To solve for the initial data
$(\overline{g}_0,k_0,\rho_0,\phi_0)$, the evolution system
$(H_{\overline{g}},H_\Phi,H_\rho)$ in the domain
$\widehat{Y}:=Y\times \mathbb{R}^n$, we first proceed to the unique
determination of the restrictions to the initial hypersurfaces
$\mathcal{I}^0$ and $\mathcal{I}^1$ of the derivatives of all order
of the possible $\mathcal{C}^\infty$ solution
$(\overline{g},\Phi,\rho)$. Then, by using some variants of the
 Borel's classical lemma and some arguments of domain of dependence,
we can transform the evolution problem into a spacelike Cauchy
problem $(\mathcal{P})$ for a third order hyperbolic system of
unknown $(h,f,A)$ defined in the domain $\widehat{\Omega}_T\equiv
\Omega_T\times \mathbb{R}^n$, with zero initial data on the
spacelike hypersurface $\Lambda_0$, where for $T>0,\;0\leq t\leq T$,
\begin{equation}\label{a4}
\Omega_t:=\{(x^\alpha)\in \mathbb{R}^{n+1}/ 0< x^0<t;|x^i|<
K(2t-x^0);\;i=1,2,...,n\} ,\;\Lambda_\tau:=\Omega_t\cap
\{x^0=\tau\},\;0\leq\tau\leq t,
\end{equation}
with $K>1$ large enough s.t. the hypersurfaces
\begin{eqnarray}
  \mathcal{H}^i &:=& \{(x^\alpha)\in \mathbb{R}^{n+1} /\;|x^i|=K(2t-x^0),\;0\leq x^0\leq t,\;i=1,2,...,n\}
\end{eqnarray}are spacelike w.r.t. the constructed metric
\begin{equation}
g_0=-\frac{|\overline{g}_0|}{|\gamma|}(dx^0)^2+\overline{g}_{0ij}dx^idx^j.
\end{equation}
 The linearized $\mathcal{C}^\infty$ problem $(\mathcal{P}_l)$
associated to $(\mathcal{P})$ is solved by applying the Leray's
theory of hyperbolic systems and the classical
method of characteristics.\\
Let $s$ the smallest integer s.t. $s>\frac{n}{2}+2$, there exists a
suitable weighted Sobolev space $\mathcal{E}^s(\widehat{\Omega}_T)$
of order $s$, in which one can develop a fixed point method based on
some energy estimates established for the $\mathcal{C}^\infty$
solution of the linearized problem $(\mathcal{P}_l)$, and which
leads to a unique solution $(h,f,A)\in
\mathcal{E}^s(\widehat{\Omega}_T)$ of the problem $(\mathcal{P})$,
for $T>0$ small enough. Then the $\mathcal{C}^\infty$ regularity of
this solution $(h,f,A)$ is established by showing by induction on
$m$ and some classical arguments \cite{o}, that $(h,f,A)\in
\mathcal{E}^m(\widehat{\Omega}_T)$ for every $m\geq s$, and finally
by using the Sobolev embedding theorem. The evolution system
$(H_{\overline{g}},H_\rho,H_\Phi)$ thus has a unique
$\mathcal{C}^\infty$ solution $(\overline{g},\rho,\Phi)$ for the
deduced initial data
$(\overline{g}_0,k_0,\rho_0,\phi_0)\blacksquare$

\begin{appendices}
\section{Sketch of the proof of theorem \ref{th3}\label{ap}}
Given the full initial data $(\overline{g}_0,k_0,\rho_0,\phi_0)$ as
constructed, in section \ref{s4},
 as the solution of the initial data constraints's problem, we consider now the characteristic Cauchy
problem
\begin{equation}\label{e2}
\mathcal{P}\left\{
  \begin{array}{ll}
  H_{\overline{g}}:\;\partial_0
R_{ij}-\overline{\nabla}_iR_{j0}-\overline{\nabla}_jR_{i0}=\partial_0
\Lambda_{ij}-\overline{\nabla}_i\Lambda_{j0}-\overline{\nabla}_j\Lambda_{i0},& \hbox{in\;$Y$,} \\
    H_\rho:\;p^\alpha\frac{\partial\rho}{\partial x^\alpha}-
    \Gamma_{\mu\nu}^i p^\mu p^\nu\frac{\partial\rho}{\partial p^i}
    =0, & \hbox{in\;$ \mathbb{P}$,} \\
       H_\Phi:\;  \square_g \Phi=V'(\Phi),
 & \hbox{in\;$Y$,} \\
(\overline{g},\partial_0 \overline{g},\rho,\Phi)_{|\mathcal{I}}=
(\overline{g}_0,k_0,\rho_0,\phi_0). & \hbox{}
  \end{array}
\right.
\end{equation}A first step towards the solving of the problem $\mathcal{P}$ consists
to determine uniquely, by induction on $k\in \mathbb{N}$, the
functions $\psi^{(k)}$ defined as the trace on the initial
hypersurface $\mathcal{I}$ of the derivatives
$\left(\frac{\partial^k \overline{g}}{(\partial x^0)^k},
\frac{\partial^k \rho}{(\partial x^0)^k},\frac{\partial^k
\Phi}{(\partial x^0)^k}\right)$ of
 the possible $\mathcal{C}^\infty$
solution $(\overline{g},\rho,\Phi)$ of the problem $\mathcal{P}$.
Using subsequently some variants of Borel's classical lemma, we can
construct an auxiliary function
 $w=(\mu=(\mu_{ij}),\varrho,\kappa)\in \mathcal{C}^\infty (\Omega_T\times \mathbb{R}^n)$
 s.t. $\left(\frac{\partial^k \mu}{(\partial x^0)^k},
\frac{\partial^k \varrho}{(\partial x^0)^k},\frac{\partial^k
\kappa}{(\partial x^0)^k}\right)_{/\mathcal{I}}=\psi^{(k)}$ for
every $k\in \mathbb{N}$, where,  for $0\leq t\leq T,\;0\leq\tau\leq
t,\;\Omega_t,\; \Lambda_\tau$ are defined in (\ref{a4}).
 The function $w$ verifies thus on $\mathcal{I}$ the evolution
system $(H_{\overline{g}},H_\rho,H_\Phi)$
 and its derivatives of all orders. Introducing now the new unknown
$v=u-w=(\widetilde{h},\widetilde{A},\widetilde{f})$ with
 $\widetilde{h}=\overline{g}-\mu,\widetilde{f}=\rho-\varrho, \widetilde{A}=\Phi-\kappa$,
 we can transform the problem $\mathcal{P}$ into a zero initial data characteristic Cauchy problem $\mathcal{P}_1$
 in $\mathbb{P}$, which is of
 the form
  \begin{equation}
   \mathcal{P}_1 \left\{%
\begin{array}{ll}
   H_{\widetilde{h}}\;:\;\widetilde{g}^{\lambda\nu}D_{\lambda\nu} \partial_0 \widetilde{h}_{ij}=
    \widetilde{F}_{ij}(x,D^\alpha \widetilde{h}_{lk},D^\beta \widetilde{f},D^\alpha \widetilde{A}),\;
    |\alpha|\leq 2,
  & \hbox{} \\
   H_{\widetilde{f}}\;:\;\widetilde{p}^\alpha\frac{\partial \widetilde{f}}{\partial x^\alpha}
    +\widetilde{P}^i\frac{\partial \widetilde{f}}{\partial p^i}
    +\mathcal{L}_Z\varrho=0,\;Z=(\widetilde{p}^\nu,\widetilde{P}^i) & \hbox{} \\
     H_{\widetilde{A}}:\; \widetilde{g}^{\lambda\nu}D_{\lambda\nu}  \widetilde{A}=\widetilde{F}(x,
,D^\beta \widetilde{h}_{lk},D^\gamma \widetilde{A})
     ,\;|\beta|\leq 1,\;|\gamma|\leq 1 & \hbox{} \\
    (\partial^k_0\widetilde{h},\partial^k_0\widetilde{f},\partial^k_0\widetilde{A})_{/\mathcal{I}}=
    (0),\;\forall\; k\in\mathbb{N}; & \hbox{} \\
         \end{array}%
\right.
\end{equation}
where $\alpha=(\alpha_0,\alpha_1,...,\alpha_n)\in
\mathbb{N}^{n+1},\;|\alpha|=\alpha_0+\alpha_1+...+\alpha_n,\;D^\alpha
\equiv \frac{\partial^{|\alpha|}}{(\partial x^0)^{\alpha_0}
      (\partial x^1)^{\alpha_1}...(\partial x^n)^{\alpha_n}}$,
      $\mathcal{L}_Z$ denotes the Lie-derivative w.r.t.
      $Z\equiv(\widetilde{p}^\nu,\widetilde{P}^l)$ with
    \begin{eqnarray*}
    \widetilde{p}^0&=&\sqrt{|\gamma|}\frac{\sqrt{\textbf{m}^2+
    \widetilde{g}_{ij}p^ip^j}}{\sqrt{|\overline{\widetilde{g}}|}},\widetilde{p}^i=p^i,
    \widetilde{P}^l=-\widetilde{\Gamma}^l_{\lambda\nu}p^\lambda p^\nu,\\
\widetilde{g}_{ij}&=&\widetilde{h}_{ij}+\mu_{ij},\;
    \overline{\widetilde{g}}=(\widetilde{g}_{ij}),\;\widetilde{g}_{00}=
    -\frac{|\overline{\widetilde{g}}|}{|\gamma|},\;\gamma=(\gamma_{ab}), a,b=2,...,n.
    \end{eqnarray*}The
    characteristic Cauchy problem $\mathcal{P}_1$ is then extended to an
    ordinary spacelike Cauchy problem $(\mathcal{P})$ in
$ \Omega_T\times \mathbb{R}^n$ with zero initial data on the the
spacelike hypersurface $\Lambda_0$, ie.:
\begin{equation}
   (\mathcal{P})\left\{%
\begin{array}{ll}
    H_{h}\;:\;\widetilde{g}^{\lambda\nu}D_{\lambda\nu} \partial_0 h_{ij}=
    \widetilde{F}_{ij}+ r_{ij}
  & \hbox{in $\Omega_T$} \\
    H_f\;:\;\widetilde{p}^\alpha\frac{\partial f}{\partial x^\alpha}
    +\widetilde{P}^i\frac{\partial f}{\partial \widetilde{p}^i}
    +\mathcal{L}_Z\varrho=r & \hbox{in $\Omega_T\times \mathbb{R}^n $} \\
     H_A:\;  \widetilde{g}^{\lambda\nu}D_{\lambda\nu} A=\widetilde{F}+r_\Phi
      & \hbox{in $\Omega_T$} \\
    ({\partial^q_0 h},\;f,\;A,\;
    {\partial_0A})_{/\Lambda_0}=0,\;q\in\{0,1,2\}).
     & \hbox{} \\
         \end{array}%
\right.
\end{equation}
 The unknowns functions are $h_{ij},A,f$;
  the $\mathcal{C}^\infty $ functions $r_{ij},r_\Phi,r$
are defined by
\begin{equation}
    r_{ij}=\left\{%
\begin{array}{ll}
    0,\;\hbox{if}\;(x^\alpha)\in Y & \hbox{} \\
    -\widetilde{F}_{ij}(x^\alpha,0,...,0),\;\hbox{if}\;
    (x^\alpha)\in \Omega_T-Y, & \hbox{} \\
\end{array}%
\right.
r_\Phi=\left\{%
\begin{array}{ll}
    0,\;\hbox{if}\;(x^\alpha)\in Y & \hbox{} \\
   - \widetilde{F}(x^\alpha,0,...,0),\;\hbox{if}\;
    (x^\alpha)\in \Omega_T-Y, & \hbox{} \\
\end{array}%
\right.
\end{equation}
\begin{equation}
  r=  \left\{%
\begin{array}{ll}
    0,\;\hbox{if}\; (x^\alpha,p^\delta)\in \mathbb{P} & \hbox{} \\
    \mathcal{L}_{Z_0} \varrho(x^\alpha,0,...,0),\;\hbox{if}\;
    (x^\alpha)\in (\Omega_T\times \mathbb{R}^n)-  \mathbb{P}, & \hbox{} \\
\end{array}%
\right.
\end{equation}where $Z_0$ is deduced of $Z$ by imposing $\widetilde{h}=0$.\\
Now, for given $\mathcal{C}^\infty$ functions
$\widehat{h}=(\widehat{h}_{ij}),\;\widehat{A}$ on $\Omega_T$,
$\widehat{f}$ on $\Omega_T\times \mathbb{R}^n$ s.t.
$\overline{\upsilon}=(\upsilon_{ij}=\mu_{ij}+\widehat{h}_{ij})$ is
properly riemannian on $\Omega_T$, and for
\begin{equation*}
    \upsilon=-\frac{|\overline{\upsilon}|}{|\gamma|}(dx^0)^2+\upsilon_{ij}dx^i
    dx^j
\end{equation*}we consider the linear problem
\begin{equation}
   (\mathcal{P}_l)\left\{%
\begin{array}{ll}
    H^l_{h}\;:\;
\upsilon^{\lambda\nu}(x^\alpha)D_{\lambda\nu}\partial_0 h_{ij}=
    \widetilde{F}_{ij}+ r_{ij}\equiv G_{ij}
  & \hbox{in $\Omega_T$} \\
     H^l_f\;:\;\sqrt{|\gamma}|\frac{\sqrt{\textbf{m}^2+
    \upsilon_{ij}p^ip^j}}{\sqrt{|\overline{\upsilon}|}}
    \;\frac{\partial f}{\partial x^0}+p^i\frac{\partial f}{\partial
    x^i}
    +\widetilde{P}^i\frac{\partial f}{\partial p^i}=
    -\mathcal{L}_Z\varrho +r  &\hbox{in\; $\Omega_T\times \mathbb{R}^n$} \\
     H^l_A:\;  \upsilon^{\lambda\nu}(x^\alpha)D_{\lambda\nu}
     A=\widetilde{F}+r_\Phi\equiv J
      & \hbox{in $\Omega_T$} \\
    ({\partial^q_0 h},\;f,\;A,\;
    {\partial_0A})_{/\Lambda_0}=0,\;q\in\{0,1,2\}).
     & \hbox{} \\
         \end{array}%
\right.
\end{equation}where the $\mathcal{C}^\infty$ functions $\widetilde{P}^i,\;
G\equiv (G_{ij}= \widetilde{F}_{ij}+
r_{ij}),\;J\equiv\widetilde{F}+r_\Phi,\;
        -\mathcal{L}_Z\varrho +r $ are taken for the metric
    $\upsilon=(\upsilon_{00},\overline{\upsilon}=(\upsilon_{ij}))$, the function $\widehat{A}$ , and the function
    $\widehat{f}$ which is supposed of compact support. Thanks to the Leray theory
of hyperbolic systems applied for the equations $H_h^l, H_A^l$, and
the classical method of characteristics applied for the equation
$H_f^l$, one has:
 {\lemma{The hypotheses are those of the theorem,
then, for given $\mathcal{C}^\infty$ functions
$\widehat{h}=(\widehat{h}_{ij}),\;\widehat{A}$ on $\Omega_T$,
$\widehat{f}$ on $\Omega_T\times \mathbb{R}^n$, the linear problem
$(\mathcal{P}_l)$ has a unique $\mathcal{C}^\infty$ solution defined
on $\Omega_T\times
\mathbb{R}^n$, with support contained in $Y\times \mathbb{R}^n$. }}\\
 Now,
 to obtain a $\mathcal{C}^\infty$ solution for the nonlinear
problem $(\mathcal{P})$, we consider a framework of weighted Sobolev
spaces in order to apply a fixed point method. Given $T>0$ and $K>1$
large enough as indicated above, considering $\Omega_t,\;\Lambda_t$
as defined in (\ref{a4}), one sets for $0\leq t\leq T$:
\begin{eqnarray}
  P_t &:=&\{(x^\alpha,p^\delta)\in \Omega_t\times \mathbb{R}^{n+1}/\upsilon_{\mu\nu}p^\mu
p^\nu=-\textbf{m}^2\}.
\end{eqnarray}One denotes by $C_0^\infty(\Omega_t)$
the space of restrictions to $\Omega_t$ of $\mathcal{C}^\infty$
functions defined in a neighborhood of $\Omega_t$ and by
  $C_0^\infty(P_t)$ the space of restrictions to $P_t$ of $\mathcal{C}^\infty$
functions with compact support in a neighborhood of $P_t$.\\
  Let $s\in \mathbb{N}$. For a
 function $v=(v_I)$
               defined in a neighborhood of  $\Omega_T$, for a function
               $f$ defined in a neighborhood of $P_T$, we introduce the norms
                       \begin{equation*}
            \|v\|^2_{H^s(\Lambda_t)}:=\sum_I\sum_{|\alpha|\leq
            s}\int_{\Lambda_t}|D^\alpha v_I|^2dx
            ,\;\|f\|^2_{F^s(\Lambda_t\times \mathbb{R}^n)}:=\sum_{|\alpha|+|\beta|\leq
            s}\int_{\Lambda_t\times \mathbb{R}^n}(p^0)^{2(|\alpha|+|\beta|)+1}|D_x^\alpha
            \partial_p^\beta f|^2dx'dp;
            \end{equation*}where for $ \alpha \in \mathbb{N}^{n+1},\;\beta\in\mathbb{N}^n$,
\begin{eqnarray*}
      D_x^\alpha\partial^\beta_p
&:=&\frac{\partial^{|\alpha|+|\beta|}}{(\partial x^0)^{\alpha_0}
      (\partial x^1)^{\alpha_1}...(\partial x^n)^{\alpha_n}(\partial p^1)^{\beta_1}
      (\partial p^2)^{\beta_2}...(\partial p^n)^{\beta_n}},\\
    dx'&:=&dx^1...dx^n,\;dp:=dp^1...dp^n,\;dx=dx^0dx';
\end{eqnarray*}and denote by
         $E^s(\Omega_t)$ (respectively $E^{s}(P_t)$ ) the closure of
        $C_0^\infty(\Omega_t)$ (respectively  $C_0^\infty(P_t)$) w.r.t. the
        norms
        \begin{equation*}
            \|v\|_{E^s(\Omega_t)}:=\sup_{0\leq \tau\leq
            t}\|v\|_{H^s(\Lambda_\tau)},\; \|f\|_{E^{s}(P_t)}:=\sup_{0\leq \tau\leq t}
 \|f\|_{F^s(\Lambda_\tau\times \mathbb{R}^n)}.
\end{equation*}
     {\lemma{For every solution $(h,f,A)\in
\mathcal{C}_0^\infty(\Omega_T)\times \mathcal{C}^\infty_0
(\Omega_{T}\times \mathbb{R}^n)\times\mathcal{C}_0^\infty(\Omega_T)$
of
       the linearized problem $(\mathcal{P}_l)$,
       one has for $s>\frac{n}{2}+2,\;t\in ]0,T]$:
       \begin{eqnarray*}
\|f\|_{E^{s}(P_t)}&\leq&
C(T)\|-\mathcal{L}_Z\varrho+r\|_{E^{s}(P_t)}.\;t;\\
\|h\|^2_{H^s{(\Lambda_t)}}&\leq& \int_0^t\left\{R_1(\tau)\|
h\|_{H^s(\Lambda_\tau)}+R_2(\tau)\|
h\|^2_{H^s(\Lambda_\tau)}\right\}d\tau,\;\;
\|h\|_{E^s{(\Omega_t)}}\leq Z_2 \|G\|_{E^{s-2}(\Omega_t)}.t;\\
\|A\|^2_{H^s{(\Lambda_t)}}&\leq& \int_0^t\left\{R_3(\tau)\|
h\|_{H^s(\Lambda_\tau)}+R_4(\tau)\|
h\|^2_{H^s(\Lambda_\tau)}\right\}d\tau,\;\;\|A\|_{E^s{(\Omega_t)}}\leq
Z_3\|J\|_{E^{s-1}(\Omega_t)}. \;t;
\end{eqnarray*}
      $Z_2,\;Z_3$ are constants depending only of $T$ and some
intrinsic constants\textbf{.} }\label{l1}}
 \\Now we consider the map
\begin{eqnarray*}
  {\L}:\left(\mathcal{C}_0^\infty(\Omega_T)\right)^{\frac{n(n+1)}{2}}\times
\mathcal{C}^\infty_0 (\Omega_{T}\times
\mathbb{R}^n)\times\mathcal{C}_0^\infty(\Omega_T) & \rightarrow &
(\mathcal{C}^\infty_0(\Omega_T))^{\frac{n(n+1)}{2}}\times
\mathcal{C}^\infty_0 (\Omega_{T}\times \mathbb{R}^n)\times\mathcal{C}_0^\infty(\Omega_T) \\
  (\widehat{h}=(\widehat{h}_{ij}),\widehat{f},\widehat{A})
  &\mapsto& (h,f,A)
\end{eqnarray*}where $(h,f,A)$ is the unique solution of the linearized
problem  $(\mathcal{P}_l)$ associated to $(\mathcal{P})$ for the
given functions
$(\widehat{h}=(\widehat{h}_{ij}),\widehat{f},\widehat{A})$; let $s$
the smallest integer such that $s> \frac{n}{2}+2$; set:
$\mathcal{E}^s\equiv\left(E^s(\Omega_T)\right)^{\frac{n(n+1)}{2}}\times
E^{s}(\Omega_T\times \mathbb{R}^n)\times E^s(\Omega_T)$. Therefore
${\L}$ extends to a map ${\L}'$ defined similarly and mapping
$\mathcal{E}^s$ into itself:
\begin{eqnarray*}
  {\L}':\mathcal{E}^s\equiv\left(E^s(\Omega_T)\right)^{\frac{n(n+1)}{2}}\times
E^{s}(\Omega_T\times \mathbb{R}^n)\times E^s(\Omega_T) & \rightarrow
&\mathcal{E}^s\equiv (E^s(\Omega_T))^{\frac{n(n+1)}{2}}\times
E^{s}(\Omega_T\times \mathbb{R}^n)\times E^s(\Omega_T) \\
  (\widehat{h}=(\widehat{h}_{ij}),\widehat{f},\widehat{A})
  &\mapsto& (h,f,A).
\end{eqnarray*}Using the energy estimates above, one shows
that there exists $T_*>0$
 small enough and  $R>0$ large enough s.t. ${\L}'$ is a contraction map
 from the closed ball $B(0,R)$ of the Banach space  $\mathcal{E}^s$ into itself.
${\L}'$ admits a fixed point which is the desired solution of
$(\mathcal{P})$. The $\mathcal{C}^\infty$ regularity of this
solution $(h,f,A)$ is established by showing by induction on $m$ and
some classical arguments \cite{o}, that $(h,f,A)\in
\mathcal{E}^m(\Omega_T\times \mathbb{R}^n)$ for every $m\geq s$, and
finally by using the Sobolev embedding theorem. The support of this
solution is
 in the domain above
$\mathcal{I}$. The evolution system
$(H_{\overline{g}},H_\Phi,H_\rho)$ thus has a unique
$\mathcal{C}^\infty$ solution $(\overline{g};\Phi,\rho)$ for the
given $\mathcal{C}^\infty$ initial data $\blacksquare$
\end{appendices}{}

Marcel Dossa\\
University of Yaounde I, Faculty of Sciences. P. O. Box. 812,
Yaounde, Cameroon; Department of Mathematics. E-mail address:
marceldossa@yahoo.fr

Jean Baptiste Patenou\\
University of Dschang, Faculty of Sciences. P. O. Box. 67 Dschang,
Cameroon; Department of Mathematics and Computer science. E-mail
address: jeanbaptiste.patenou@univ-dschang.org

\end{document}